\documentstyle[12pt,moriond,epsfig]{article}

\newcommand{\bgeq}{\begin{equation}}
\newcommand{\edeq}{\end{equation}}
\newcommand{\bgea}{\begin{eqnarray}}
\newcommand{\edea}{\end{eqnarray}}
\begin{document}
\vspace*{2cm}
\title{TRIAXIAL NEUTRON STARS --- A POSSIBLE \\ SOURCE OF GRAVITATIONAL
RADIATION}
\author{Silvano Bonazzola, Joachim Frieben, and Eric Gourgoulhon}
\affil{D\'epartement d'Astrophysique Relativiste et de Cosmologie,
Unit\'e Propre 176 du {\sf CNRS},
Observatoire de Paris,
F--92195 Meudon Cedex, France}
\vspace*{-0.1cm}
\begin{figure}[h]\label{FIG:Author}
\unitlength1cm
\begin{center}
\begin{picture}(4,5)
\epsfig{file=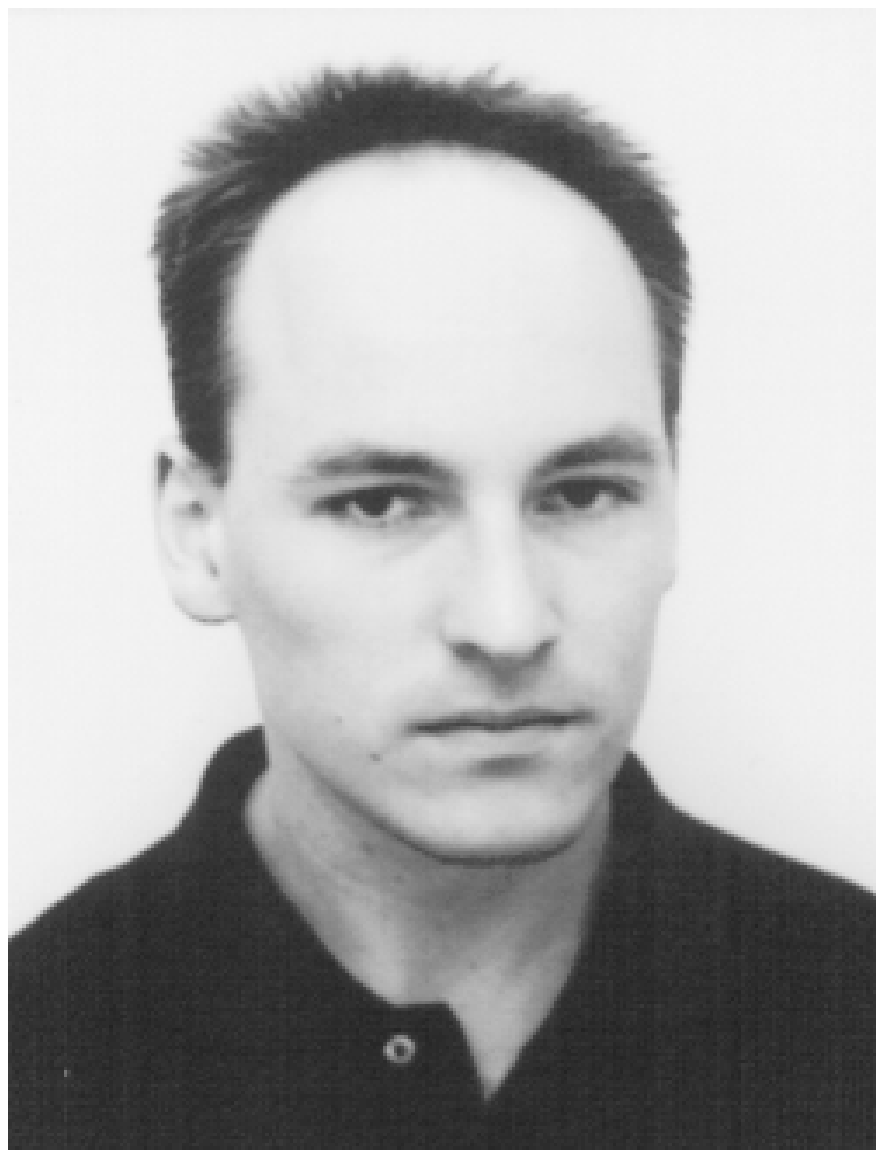,height=5cm}
\end{picture}
\end{center}
\end{figure}
\begin{abstract}
Triaxial neutron stars may be important sources of gravitational radiation
for the forthcoming generation of interferometric gravitational wave detectors
such as {\sf LIGO}, {\sf VIRGO}, and {\sf GEO600}. We investigate the
viscosity triggered bar mode secular instability of rapidly rotating neutron
stars by means of a perturbation analysis of numerically constructed {\em
``exact''\/} general relativistic axisymmetric star models.
In the theoretical approach, only the dominant parts of the nonaxisymmetric
terms of the 3D--Einstein equations are taken into account. A comparison of
our results with previous studies of {\em Newtonian\/} polytropic stars
confirms James' classical result $\gamma_{\rm crit}\!=\!2.238$ for the critical
polytropic index. Beyond the Newtonian regime, $\gamma_{\rm crit}$ reveals a
slight increase toward highly relativistic configurations. Six out of twelve
employed realistic dense matter equations of state admit the spontaneous
symmetry breaking for masses above $1.6\,M_\odot$.
\renewcommand{\baselinestretch}{0.85}\normalsize
\end{abstract}
\vspace*{-1ex}
\section{Introduction}\vspace*{-4mm}
Rapidly rotating neutron stars are highly relativistic objects, and, provided
there is some physical process operative, which induces a significant deviation
from axisymmetry, may be important sources of gravitational radiation.
A transition toward triaxial configurations can occur, when the ratio $T/|W|$
of rotational kinetic energy and gravitational potential energy reaches some
critical value$^{1,2)}$. The equation of state, hereafter EOS, has an
important influence on the development of a triaxial instability since the
neutron star matter must be stiff enough to admit a maximum angular velocity
$\Omega_{\rm K}$ higher than the critical one. Homogeneous, incompressible
fluid bodies, rotating at moderate constant angular velocity, take the shape
of some oblate axisymmetric {\em Maclaurin spheroid\/}. Maclaurin
spheroids are dynamically unstable for $T/|W|\!>\!0.2738$, but there are two
other families of triaxial ellipsoids which bifurcate from the Maclaurin
sequence earlier at $T/|W|\!=\!0.1375$. We focus on the sequence of {\em
Jacobi ellipsoids\/} which rotate uniformly about their smallest axis in an
inertial frame. The evolution toward a Jacobi ellipsoid at the bifurcation
point is triggered by some {\em viscous\/} dissipative mechanism. This
instability is called a {\em secular\/} one, because it evolves on the
associated viscous timescale which is much longer than the dynamical one.
The astrophysical realization corresponds to the scenario of a cold highly
viscous binary neutron star, being spun up by accretion from its companion.
Viscosity dissipates mechanical energy, while it preserves angular momentum.
As a consequence, the Maclaurin spheroid, once the instability point is reached,
develops toward a Jacobi ellipsoid which exhibits the lowest rotational kinetic
energy for a fixed value of the angular momentum. In this final state, viscous
dissipation has ceased, and the star is rotating rigidly. Former studies had
been performed at the Newtonian$^{3,4,5,6)}$ or Post--Newtonian$^{7,8)}$
level which is by no means adequate for highly relativistic objects like
neutron stars. Furthermore, the neutron star matter was modeled by a
simplifying polytropic equation of state.
\vspace*{-1ex}
\section{Numerical models of triaxial neutron stars}\vspace*{-4mm}
\begin{figure}[t]
\vspace*{-8mm}
\parbox{90mm}{\epsfig{figure=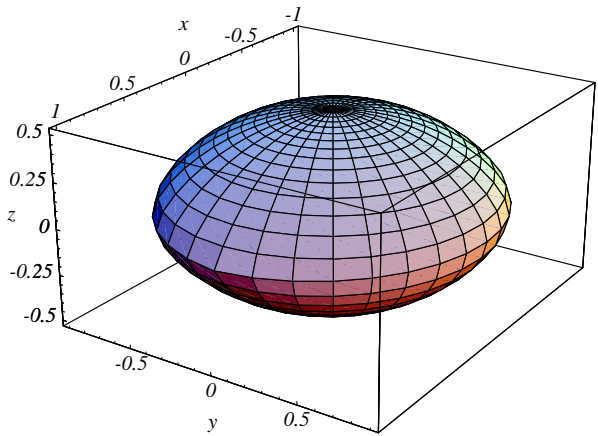,angle=270,width=80mm}}
\parbox{70mm}{
\caption[]{Triaxial maximum rotation New\-ton\-ian polytropic star for
	$\gamma\!=\!2.4$. The ellipticity in the equatorial plane is
	about 0.1. Notice the cusp at the stellar equator in direction
	of the semimajor axis ($x$--axis), where the rotation is Keplerian,
	and which is absent along the semiminor axis ($y$--axis).
	\label{FIG:nsmax}}}
\vspace*{-10mm}
\end{figure}
Before the symmetry breaking sets in, the neutron star, modeled as a rigidly
rotating perfect fluid, can be considered as {\em stationary\/} and {\em
axisymmetric\/}. Making the further assumption that no {\em meridional\/} matter
currents are present, a favourable choice of {\em elliptic\/} field equations
was given by Bonazzola et al.$^{9)}$. The numerical code$^{9)}$, based upon
these equations, relies on a spectral method, and allows to compute neutron
star models with a precision of $\simeq\!10^{-14}$ in the spherical symmetric
case and $\simeq\!10^{-6}$ for maximum rotation configurations, when a
$\gamma\!=\!2$ polytropic, {\em analytic\/} EOS is used, as well as
$\simeq\!10^{-4}$ for realistic EOS$^{10)}$.
The neutron star models are {\em ``exact''\/} in the sense that the full
Einstein equations are solved without any analytic approximation, while the
numerical integration covers all space and respects the exact flat space
boundary condition at spatial infinity.
When the symmetry breaking occurs, spacetime is neither stationary nor
axisymmetric. However, at the very beginning, the deviation from axisymmetry is
sufficiently small, and the emitted gravitational radiation may be neglected.
Under the additional assumption of {\em rigid\/} rotation, the {\em helical\/}
symmetry of spacetime is preserved. Taking into account only the dominant
non--axisymmetric terms, one basically recovers the original equations of the
axisymmetric case$^{11)}$. The only difference is that the lapse function $N$
--- its logarithm $\nu\!=\!\log N$ reduces to the Newtonian potential in the
weak field approximation --- and the matter fields are three--dimensional
quantities which depend on $r$, $\theta$, and $\psi\equiv\phi-\Omega\,t$, where
$\Omega$ is the angular velocity of the star. As a consequence, the present
approach is exact for the stationary and axisymmetric relativistic, as well as
for the fully three--dimensional Newtonian case. After relaxation to a
particular axisymmetric configuration, a small perturbation
$\delta\nu\!=\!\epsilon H_{\rm c}(r\sin\theta\cos\psi)^2$, which excites the
$l\!=\!2$, $m\!=\!\pm 2$ bar mode, is added to $\nu\!=\!\log N$, and the growth
of $\delta\nu$ during the subsequent relaxation is followed. $H_{\rm c}$
denotes the central log--enthalpy, and $\epsilon$ is a small parameter of the
order of $10^{-6}$.
A particular configuration is {\em secularly unstable\/} if the perturbation
increases, and the fluid body evolves subsequently toward a triaxial spheroid.
It is {\em conditionally secularly stable\/} if the perturbation tends to zero.
In this case, however, three--dimensional terms of higher relativistic order,
which have not been taken into account at the present level of approximation,
may induce the symmetry breaking.
\vspace*{-1ex}
\section{Results for polytropic and realistic equations of state}\vspace*{-4mm}
\begin{figure}[t]
\vskip-5mm
\parbox{100mm}{\epsfig{figure=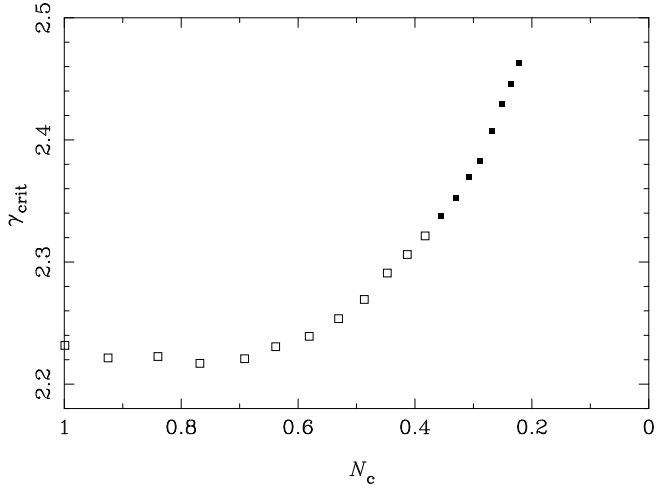,angle=270,width=100mm}}
\parbox{60mm}{
\caption[]{Critical polytropic index $\gamma_{\rm crit}$ as function of the
	lapse function $N_{\rm c}$, measured at the centre of the star. Black
	boxes indicate configurations unstable with respect to radial
	oscillations.
	\label{FIG:gcrit-Nc}}}
\end{figure}
\begin{table}[t]
\begin{center}
\parbox{110mm}{
\begin{small}
\begin{tabular}{lllllll}
  $\displaystyle{{\rm EOS}\atop \ }$ &  
  $\displaystyle{{M_{\rm max}^{\rm stat}\atop [M_\odot]}}$ &
  $\displaystyle{{M_{\rm max}^{\rm rot}\atop [M_\odot]}}$ &
  $\displaystyle{{P_{\rm K}\atop [{\rm ms}]}}$ &
  $\displaystyle{{P_{\rm break}\atop [{\rm ms}]}}$ &
  $\displaystyle{{H_{\rm c,break}\atop \ }}$ &
  $\displaystyle{{M_{\rm break}\atop [M_\odot]}}$ 	\\[1.5ex]
\tableline
  HKP	  & 2.827 & 3.432 & 0.737 & 1.193 & 0.168 & 1.886 \\
  WFF2	  & 2.187 & 2.586 & 0.505 & 0.764 & 0.292 & 1.925 \\
  WFF1 	  & 2.123 & 2.528 & 0.476 & 0.728 & 0.270 & 1.742 \\
  WGW  	  & 1.967 & 2.358 & 0.676 & 1.042 & 0.170 & 1.645 \\
  Glend3  & 1.964 & 2.308 & 0.710 & \multicolumn{3}{c}{stable} \\
  FP	  & 1.960 & 2.314 & 0.508 & 0.630 & 0.412 & 2.028 \\
  DiazII  & 1.928 & 2.256 & 0.673 & \multicolumn{3}{c}{stable} \\
  BJI	  & 1.850 & 2.146 & 0.589 & \multicolumn{3}{c}{stable} \\
  WFF3    & 1.836 & 2.172 & 0.550 & 0.712 & 0.327 & 1.919 \\
  Glend1  & 1.803 & 2.125 & 0.726 & \multicolumn{3}{c}{stable} \\
  Glend2  & 1.777 & 2.087 & 0.758 & \multicolumn{3}{c}{stable} \\
  PandN   & 1.657 & 1.928 & 0.489 & \multicolumn{3}{c}{stable}
\end{tabular}
\end{small}}
\quad
\parbox{50mm}{
\caption[]{Neutron star properties according to various EOS: $M_{\rm max}^{\rm
	stat}$ is the maximum mass for static configurations, $M_{\rm max}^{\rm
	rot}$ is the maximum mass for rotating stationary configurations,
	$P_{\rm K}$ is the corresponding Keplerian period, $P_{\rm break}$ is
	the rotation period below which the symmetry breaking occurs, $H_{\rm
	c, break}$ is the central log--enthalpy at the bifurcation point and
	$M_{\rm break}$ is the corresponding gravitational mass. The EOS are
	ordered by decreasing values of $M_{\rm max}^{\rm stat}$.}
	\label{TAB:res-EOS}}
\end{center}
\vskip-5mm
\end{table}
\begin{figure}[b]
\vskip-15mm
\hspace{-10mm}
\parbox{110mm}{\epsfig{figure=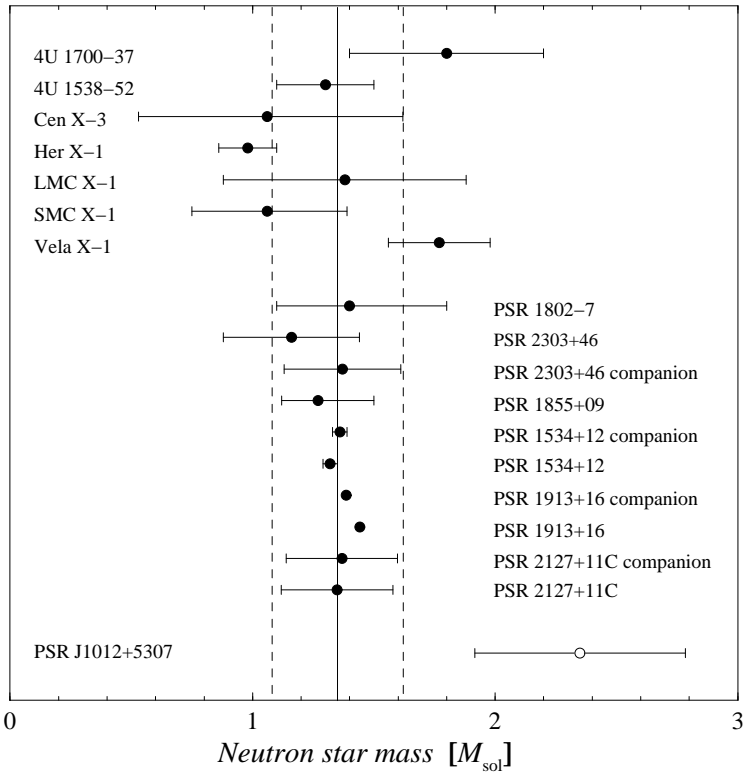,angle=270,width=130mm}}
\hspace*{15mm}
\parbox{50mm}{
\caption[]{Measured masses of 18 neutron stars. Objects in massive X--ray
	binaries are at the top, radio pulsars and their companions at the
	bottom. Data compiled by Thorsett et al.$^{12)}$ is marked by filled
	circles. The empty circle indicates the mass of the recently discovered
	radio pulsar PSR J1012+5307, measured by van Kerkwijk et al.$^{13)}$.
	The continuous line indicates the estimate of the mean value of the
	neutron star mass distribution derived by Thorsett et al. and the
	dashed lines the corresponding $1\sigma$ interval.
	\label{FIG:nsmass}}}
\vskip-5mm
\end{figure}
The investigation of the symmetry breaking of relativistic polytropic stars
avoids the problems associated with the use of realistic dense matter
EOS$^{10)}$, and allowed us to perform some comparison with former results
obtained for Newtonian stars$^{4,5,6)}$. In particular, we have confirmed
James'$^{4)}$ value $\gamma_{\rm crit}\!=\!2.238$ of the critical polytropic
index, where secular instability sets in at maximum angular velocity, as well
as the asymptotic ratio $T/|W|_{\gamma\rightarrow\infty}\!=\!0.1375$.
No results in the fully relativistic regime had been reported so far. In the
Newtonian case, polytropic stars obey a scaling law, and the critical
polytropic index is a global constant. Relativistic effects, however, are
supposed to influence the symmetry breaking. Fig.~\ref{FIG:gcrit-Nc} shows the
dependence of $\gamma_{\rm crit}$ on the central value of the lapse function
$N_{\rm c}$ which is an appropriate signature of the relativistic character of
the star. After a very slight decrease in the weakly relativistic regime it
grows by about $10\,\%$ for strongly relativistic configurations. One concludes
furthermore that {\em any\/} neutron star, built upon a polytropic EOS and
stable against radial perturbations, becomes secularly unstable for
$\gamma\!>\!2.33$ at some critical angular velocity $\Omega_{\rm crit}\!
<\!\Omega_{\rm K}$.
The results for twelve realistic EOS of neutron star matter are presented in
Tab.~\ref{TAB:res-EOS}. A description of each EOS can be found in Salgado et
al.$^{10)}$, where they had already been employed to construct realistic
high precision neutron star models. Six equations turn out to be stable up to
the mass--shedding limit $\Omega_{\rm K}$. Only the stiffest ones allow for the
symmetry breaking. The correlation with the corresponding maximum mass is not
strict because of the density dependence of the EOS. This is notably the case
for the WGW EOS which is the stiffest one at moderate densities.
For the present sample of realistic EOS, $M_{\rm break}$ has a lower bound of
$1.645\,M_{\odot}$. This is a low value compared with maximum masses of
rotating neutron stars built upon a stiff EOS, e.g. in Tab.~\ref{TAB:res-EOS}.
Unfortunately, the real distribution of neutron star masses is still quite
uncertain. Fig.~\ref{FIG:nsmass} shows the measured masses of 17 neutron stars,
compiled by Thorsett et al.$^{12)}$, and yielding an estimate of the average
neutron star mass of $M_{\rm ns}\!=\!1.35\pm 0.27\,M_{\odot}$, as well as a
recent mass measurement of PSR J1012+5307 of $2.35\pm 0.434\,M_{\odot}{}^{13)}$.
At least the masses of the two X--ray binary neutron stars 4U 1700--37,
Vela X--1, and that of the radio pulsar PSR J1012+5307 are well above our
minimum mass of $1.645\,M_{\odot}$. If they prove to be valid after further
reduction of the observational errors, we may expect a significant number of
neutron stars susceptible of encountering the spontaneous symmetry breaking.
Their observation would provide us with useful information about the generation
of gravitational waves and constraints on real neutron star matter EOS.
\vspace*{-1ex}
\acknowledgments
J. Frieben gratefully acknowledges financial support by the {\sc Gottlieb
Daimler$-$und Karl Benz$-$Stiftung}.
\references\vspace*{-1mm}
\begin{small}
\reference
\mbox{1) } Chandrasekhar, S. 1969, Ellipsoidal figures of equilibrium,
Yale University Press, New Haven
\vskip-1.3ex
\reference 
\mbox{2) } Tassoul, J.L. 1978, Theory of rotating stars, Princeton University
Press, Princeton
\vskip-1.3ex
\reference 
\mbox{3) } Jeans, J.H. 1928, Astronomy and cosmogony, Cambridge University,
Cambridge Press\vskip-1.3ex
\reference 
\mbox{4) } James, R.A. 1964, \apj, 140, 552
\vskip-1.3ex
\reference 
\mbox{5) } Ipser, J.R., Managan, R.A. 1981, \apj, 250, 362
\vskip-1.3ex
\reference
\mbox{6) } Hachisu, I., Eriguchi, Y. 1982, Prog. Theor. Phys., 68, 206
\vskip-1.3ex
\reference
\mbox{7) } Chandrasekhar, S. 1967, \apj, 148, 621
\vskip-1.3ex
\reference
\mbox{8) } Tsirulev, A.N., Tsvetkov, V.P. 1982, \sovast, 26, 407
\vskip-1.3ex
\reference
\mbox{9) } Bonazzola, S., Gourgoulhon, E., Salgado, M., \& Marck,
J.A. 1993, \aap, 278, 421
\vskip-1.3ex
\reference
\mbox{10)} Salgado, M., Bonazzola, S., Gourgoulhon, E., \& Haensel,
P. 1994, \aap, 291, 155
\vskip-1.3ex
\reference
\mbox{11)} Bonazzola, S., Frieben, J., \& Gourgoulhon, E., 1996, \apj, 460, 379
\vskip-1.3ex
\reference 
\mbox{12)} Thorsett, S.E., Arzoumanian, Z., McKinnon, M.M., \& Taylor,
J.H. 1964, \apj, 140, 552
\vskip-1.3ex
\reference 
\mbox{13)} van Kerkwijk, M.H., Bergeron, P., \& Kulkarni,
S.R., submitted to \apjl, preprint astro--ph/9606045
\end{small}
\vspace*{4ex}
\begin{center}
{\bf ETOILES A NEUTRONS TRIAXIALES --- UNE SOURCE \\
EVENTUELLE DE RAYONNEMENT GRAVITATIONNEL}\\[2.5ex]
\end{center}
Des \'etoiles \`a neutrons triaxiales pourraient \^etre des sources de
rayonnement gravitationnel importantes pour la prochaine g\'en\'eration
de d\'etecteurs interf\'erom\'etriques d'ondes gravitationnelles comme
{\sf LIGO}, {\sf VIRGO} ou {\sf GEO600}. Nous \'etudions l'instabilit\'e
s\'eculaire du mode barre des \'etoiles \`a neutrons en rotation rapide par
l'interm\'ediaire d'une analyse perturbative de mod\`eles stellaires
axisym\'etriques {\em ``exacts''\/} en relativit\'e g\'en\'erale.
Dans l'approche th\'eorique on ne tient compte que de la partie dominante des
termes non--axisym\'etriques des \'equations d'Einstein. Une comparaison avec
des \'etudes pr\'ec\'edentes d'\'etoiles {\em newtoniennes\/} confirme le
r\'esultat classique de James \`a savoir que l'indice polytropique critique
vaut $\gamma_{\rm crit}\!=\!2.238$.
Au--del\`a du r\'egime newtonien $\gamma_{\rm crit}$ montre une faible
croissance vers les configurations fortement relativistes. Six sur douze
\'equations d'\'etat de la mati\`ere dense utilis\'ees admettent la brisure
spontan\'ee de sym\'etrie pour des masses sup\'erieures \`a $1.6\,M_\odot$.
\end{document}